\documentstyle[12pt,epsf]{article}
\newcommand{\fig}[3]{\epsfxsize=#1\epsfysize=#2\epsfbox{#3}}
\newcommand{\centre}[1]{\begin{array}{c} #1 \end{array}{}} 
\def\journal#1, #2, #3, #4 { {\sl #1~}{\bf #2~}(#3) #4 }

\def\cmp{\journal Comm. Math. Phys., }

\def\np{\journal Nucl. Phys., }

\def\pl{\journal Phys. Lett., }

\catcode`\@=11
\def\marginnote#1{}
\newcount\hour
\newcount\minute
\newtoks\amorpm
\hour=\time\divide\hour by60
\minute=\time{\multiply\hour by60 \global\advance\minute
by-\hour}\edef\standardtime{{\ifnum\hour<12
\global\amorpm={am}%
        \else\global\amorpm={pm}\advance\hour by-12 \fi
        \ifnum\hour=0 \hour=12 \fi
        \number\hour:\ifnum\minute<10
0\fi\number\minute\the\amorpm}}
\edef\militarytime{\number\hour:\ifnum\minute<10
0\fi\number\minute}

\def\draftlabel#1{{\@bsphack\if@filesw {\let\thepage\relax
   \xdef\@gtempa{\write\@auxout{\string
      \newlabel{#1}{{\@currentlabel}{\thepage}}}}}\@gtempa
   \if@nobreak \ifvmode\nobreak\fi\fi\fi\@esphack}
        \gdef\@eqnlabel{#1}}
\def\@eqnlabel{}
\def\@vacuum{}
\def\draftmarginnote#1{\marginpar{\raggedright\scriptsize\tt#1}}
\def\draft{\oddsidemargin -.5truein
        \def\@oddfoot{\sl preliminary draft \hfil
        \rm\thepage\hfil\sl\today\quad\militarytime}
        \let\@evenfoot\@oddfoot \overfullrule 3pt
        \let\label=\draftlabel
        \let\marginnote=\draftmarginnote

\def\@eqnnum{(\theequation)\rlap{\kern\marginparsep\tt\@eqnlabel}%
\global\let\@eqnlabel\@vacuum}  }

\def\numberbysection{\@addtoreset{equation}{section}
        \def\theequation{\thesection.\arabic{equation}}}

\def\underline#1{\relax\ifmmode\@@underline#1\else
 $\@@underline{\hbox{#1}}$\relax\fi}

\catcode`@=12
\relax

\numberbysection

\def\fin{\end{document}}
\def\beq{\begin{equation}}
\def\eeq{\end{equation}}

\def\beqa{\begin{eqnarray}}
\def\eeqa{\end{eqnarray}}
 \def\nnn{\nonumber \\}
\def\up#1{\begin{array}{c}\displaystyle{#1}\nnn\phantom{\displaystyle{#1}}\end{array}}

\def\pish{{\pi\over h}}
\def\hspi{{h\over \pi}}

\def\Jb{{\overline  J}}
\def\Jbhat{{\widehat{\overline J}}}
\def\me{m^e\, \! }
\def\m#1{m^e_{#1}}
\def\Je{J^e\, \!}
\def\J#1{J^e_{#1}\,\!}

\def\Jeb{{\overline J}^e\, \!}
\def\meb{{\overline m}^e\, \!}

\def\Jne#1 {J_{#1}^e\, \!}
\def\Jneb#1 {{\overline J}_{#1}^e\, \!}
\def\Jnep#1 {J_{#1}'\, \!^e\, \!}
\def\Jnebp#1 {{\overline J}_{#1}'\, \!  \!^e\, \!}

\def\Jehat{{\widehat J}^e}
\def\hhat{{\widehat h}}

\def\Jhat{{\widehat J}}

\def\mhat{{\widehat m}}

\def\chib{{\overline \chi}}

\def\nuhat{{\widehat \nu}}

\def\qhat{{\widehat q}}

\def\varpib{{\overline \varpi}}

\def\mhat{{\widehat m}}

\def\Jb{{\overline J}}
\def\mb{{\overline  m}}

\def\zb{{\bar z}}

\def\chib{{\overline \chi}}

\def\Vt{{\widetilde V}}
\def\Vbt{{{\widehat{\overline V}}}}

\def\vertex#1,#2,{{{\cal V}^{{#1,#2}}}}
\def\vertexc#1,#2,{{{\cal V}^{{#1,#2}}_{conj}}}

\def\Jgen#1 {  {\underline J_{#1}} }
\def\Jgenp#1 #2 {(J_{#1}+{#2},\Jhat_{#1})}
\def\Jgenm#1 #2 {(J_{#1}-{#2},\Jhat_{#1})}
\def\Jg#1 {J_{#1},\Jhat_{#1}}
\def\Jgp#1 #2 {J_{#1}+{#2},\Jhat_{#1}}
\def\Mgen#1 {{\underline M_{#1}}}

\def\produit#1,#2,#3,#4 {P\Bigl ( [{#1},{#2}]\otimes\{{#3}\},{#4}\Bigr )}
\def\produitscript#1,#2,#3,#4 {P\Bigl (
[{\scriptstyle{#1},{#2}}]\otimes\{{\scriptstyle{#3}}\},{#4}\Bigr
)}
\def\pprod#1,#2,#3,#4,#5 {P\Bigl ( [{#1},{#2}]\otimes[{#3},{#4}],{#5}\Bigr )}
\def\pprodscript#1,#2,#3,#4,#5 {P\Bigl (
[{\scriptstyle{#1},{#2}}]\otimes[{\scriptstyle{#3},{#4}}],{#5}\Bigr )}

\def\fusV#1,#2,#3,#4,#5,#6 {f_V(
\Jgen{#1} ,
\Jgen{#2} ,
\Jgen{#3} ,
\Jgen{#4} ,
\Jgen{#5} ,
\Jgen{#6} )}

\def\brdV#1,#2,#3,#4,#5,#6 {b_V(
\Jgen{#1} ,
\Jgen{#2} ,
\Jgen{#3} ,
\Jgen{#4} ,
\Jgen{#5} ,
\Jgen{#6} )}

\def\fusxi#1,#2,#3 {f_\xi (\Jgen{#1} ,
\Mgen{#1} ,
\Jgen{#2} ,
\Mgen{#2} ,
\Jgen{#3} )}

\def\gaghat{{\hat {\bigl \{}}}
\def\gadhat{{\hat {\bigr \}}}}

\def\bverthat{{\hat {\bigl \vert}}}

\def\sixjxi#1,#2,#3,#4,#5,#6 {{\left\{\left . \!\! \,^{#1}_{#2}
\,^{#3}_{#4} \right | \!\, ^{#5}_{#6}\right\}}}
\def\sixje#1,#2,#3,#4,#5,#6 {{\left\{\left\{\left . \!\! \, ^{#1}_{#2}
\, ^{#3}_{#4} \right | \!\, ^{#5}_{#6}\right\}\right\}}}
\def\sixjxihat#1,#2,#3,#4,#5,#6 {{{\gaghat\left . \!\! \, ^{#1}_{#2}
\, ^{#3}_{#4} \right | \!\, ^{#5}_{#6}\gadhat}}}
\def\sixjehat#1,#2,#3,#4,#5,#6 {{\gaghat\gaghat\left . \!\! \, ^{#1}_{#2}
\, ^{#3}_{#4} \right | \!\, ^{#5}_{#6}\gadhat\gadhat}}

\def\Jehat {{\widehat J^e}}
\def\me {{m^e}}

\def\mhatb{\widehat {\overline m}}
\begin{document}
\begin{titlepage}
\begin{flushright}

LPTENS--96/15, \\
hep-th/9606151, \\
January 1996
\end{flushright}

\vglue 2.5  true cm
\begin{center}
{\large \bf
Chirality Deconfinement Beyond the 
$C=1$ Barrier of \\ Two Dimensional Gravity} 
\vglue 1.5 true cm
{\bf Jean-Loup~GERVAIS}\\
\medskip
\medskip
{\footnotesize Laboratoire de Physique Th\'eorique de
l'\'Ecole Normale Sup\'erieure\footnote{Unit\'e Propre du
Centre National de la Recherche Scientifique,
associ\'ee \`a l'\'Ecole Normale Sup\'erieure et \`a
l'Universit\'e
de Paris-Sud.},\\
24 rue Lhomond, 75231 Paris CEDEX 05, ~France}.
\end{center}
\vfill
\begin{abstract}
\baselineskip .4 true cm
\noindent
{\footnotesize The characteristic novel features of strongly coupled gravity
at the special values ($C_L=7, 13, 19$) are reviewed in a
simple manner using pictures as much as possible.\\ 
(Notes of lectures at the 1995 Cargese Meeting
{\sl Low Dimensional Applications of Quantum Field Theory)} }  
\end{abstract}
\vfill
\end{titlepage}

\section{Introduction} 
In recent times we have witnessed important developments in our 
understanding of two dimensional gravity\index{Two dimensional gravity} 
in the weak coupling regime, especially in 
what concerns the dressing by gravity of matter with central charge smaller 
than one. Nevertheless, the popular approaches seem to be unable to 
overcome the so called $c=1$ barrier, powerful and elegant as they may be. 
The point of these two lectures 
 is to show, following refs.\cite{G3,GR1,GR2,GR3} how  
the operator approach to
Liouville theory\index{Liouville theory} 
 which I started to develop  long ago with A. Neveu, 
  remains  
applicable in  the strong coupling regime\index{Strong coupling regime}.
The principle is as follows. In the weak coupling regime,  the Liouville exponentials 
are expressed in terms of chiral vertex operators, themselves constructed from 
a free field in two dimensions. This is the quantum version of the well known classical 
B\"acklund transformation. Locality and closure  of their OPA 
(operator product algebra\index{Operator product algebra}) 
  uniquely determines  this chiral decomposition. In the strong coupling regime, this latter 
construction   
looses meaning since the 
OPA of the Liouville exponentials involves operators and/or
highest weight
states with
complex  Virasoro weights. Nevertheless   the general
operator-family of chiral components may still be
used, when 
truncation theorems\cite{GN5,G3,GR1} apply, that is with central charges 
$C_L=7$, $13$, $19$.  
Indeed for these values 
 there exist
subfamilies of the  above chiral operators
which form  closed operator algebras, only involving real   Virasoro weights.
In the present operator approach they   are  used to construct
a new set of local fields which replace the Liouville
exponentials. Since  both sets are constructed out of the same
free B\"acklund fields, they may be considered as  related by
a new type of quantum B\"acklund 
transformation\index{Quantum B\"acklund transformation}, that connect the
weak and strong coupling regimes of two-dimensional gravity.
 In the present lecture notes,
I summarise   the basic  features 
(\index{Deconfinement of chirality}deconfinement of chirality,
new expression for the string susceptibility\index{String susceptibility})  
of these 
new set of local fields that replace the Liouville exponentials,
in the strong coupling regime. Moreover, the basic properties of the 
new topological models\index{Topological models} 
 will be recalled, where the gravity part is in the 
strong coupling regime. The message will be that
the derivation of the new features is very close to
the previous weak-coupling  one, once the new set of local
physical fields is established. 

\section{Basic points about Liouville theory}
\subsection{The classical case}
In order to set the stage, we recall the classical structure 
for the special case of the Liouville
theory. We shall work with Euclidean coordinates
$\sigma$, $\tau$. As a preparation for the quantum case, the 
classical action is defined as 
\begin{equation}
 {\it S}={1\over 8\pi } \int d\sigma d\tau
\, \Bigl ( ({\partial \Phi\over \partial \sigma})^2
+({\partial \Phi\over \partial \tau})^2
+e^{\displaystyle 2\sqrt{\gamma} \Phi} \Bigr ).
\label{c2.1}
\end{equation}
We use world sheet variables $\sigma$ and $\tau$, which are 
local coordinates such
that the metric tensor takes the form $h_{ab} =\delta_{a,b}\,
e^{\displaystyle 2\Phi \sqrt{\gamma}}$. The
complex structure is assumed to be such that the curves  with
constant $\sigma$ and $\tau$ are everywhere
tangent to the local imaginary
and real axis  respectively.
The action \ref{c2.1}  corresponds to a conformal
 theory such that $\exp(2\sqrt{\gamma} \Phi) d\sigma d\tau $
is invariant. It is convenient to let 
\beq 
  x_{\pm }=\sigma\mp   i \tau ,
\quad
\partial_\pm = {1\over 2} ( \partial_\sigma \pm i \partial_\tau).
\label{2.2}
\eeq
By minimizing the above action, one derives the Liouville equations
\begin{equation}
 {\partial^2 \Phi\over \partial \sigma^2}
+{\partial^2 \Phi\over \partial \tau^2}=
 \sqrt{\gamma}\> e^{\displaystyle 2\sqrt{\gamma}\Phi}. 
\label{2.3}
\end{equation}
The chiral modes
 may be separated very simply using the fact that
the field $\Phi (\sigma,\, \tau)$ satisfies the above equation 
if and only if\footnote{ The factor $i$ means that these
solutions should be considered in Minkowsky space-time } 
\beq
e^{-\displaystyle \sqrt{\gamma}\Phi}={i\sqrt{\gamma} \over 2}
\sum_{j=1,2} f_j(x_+)
g_j(x_-);
  \quad  x_\pm=\sigma\mp i\tau
\label{2.4} 
\end{equation} 
where $f_j$ (resp.($g_j$), which are functions of a single variable,  are
solutions of the  same Schr\"odinger equation (primes mean derivatives) 
\beq
-f_j''+T(x_+)f_j=0,\quad
\hbox{( resp.}\>  -g''_j+ \overline T(x_-)g_j\,\hbox{)}.
\label{2.5}
\end{equation}
 The solutions
are normalized such that their Wronskians $f_1'f_2-f_1f_2'$
and $g_1'g_2-g_1g_2'$ are equal to one.
The proof goes as follows. 

 1) First check that Eq.\ref{2.4} is indeed solution. Taking the Laplacian
 of the logarithm of the right-hand side gives 
$${\partial^2 \Phi\over \partial \sigma^2}
+{\partial^2 \Phi\over \partial \tau^2}
\equiv 4\partial_+\partial_-\Phi
=-4 \sqrt{\gamma} \Bigl / \bigl( \sum_{i=1,2} f_i g_i\bigr)^2
$$
where $\partial_\pm=(\partial/\partial \sigma
\pm i\partial /\partial\tau)/2$  .
The numerator has been simplified by means of the Wronskian condition.
This is equivalent to Eq.\ref{2.3}. 

2) Conversely check that any solution of Eq.\ref{2.3} 
may be put
under the form  Eq.\ref{2.4}. If Eq.\ref{2.3} holds  
one deduces
\beq
 \partial_\mp T^{(\pm)}=0;\quad \hbox{with}\,
T^{(\pm)}:=e^{\displaystyle \sqrt{\gamma} \Phi} \partial ^2_\pm  
e^{-\displaystyle \sqrt{\gamma} \Phi}
\label{2.6}   
\end{equation}
$T^{(\pm)}$ are thus functions of a single variable. Next
the equation involving $T^{(+)}$ may be rewritten as
\beq
(-\partial_+^2 +T^{(+)}\bigr)e^{-\displaystyle \Phi}=0
\label{2.7}   
\end{equation}
with solution 
$$e^{-\displaystyle \sqrt{\gamma} \Phi}=
{i\sqrt\gamma\over 2} \sum_{j=1,2} f_j(x_+) g_j(x_-);
\quad \hbox{with}\, -f_j''+T^{(+)}f_j=0
$$
where the $g_j$ are arbitrary functions of $x_-$. Using the equation 
\ref{2.6} that involves $T^{(-)}$, one finally derives the
Schr\"odinger equation  $-g_j''+T^{(-)}g_j=0$. Thus the theorem holds
with $T=T^{(+)}$ and $\overline T=T^{(-)}$.   
One may deduce from Eq.\ref{2.6} that  the potentials of the two Schr\"odinger equations
coincide with the two chiral components of the stress-energy tensor.
 Thus these equations are the classical equivalent of the
Ward identities
that ensure the decoupling of Virasoro null vectors. 
Next a B\"acklund transformation to free fields is easily obtained as follows. It is easy 
to verify explicitly that, since $f_1$ is a solution of Eq.\ref{2.5}, as second 
solution is given by $f_2(x)=f_1(x) \int^x dy f_1^{-2}(y)$. So we may introduce a field 
$q(x)$ by letting 
\beq
f_1(x)=e^{q(x)\sqrt{\gamma}}, \quad f_2(x)=f_1(x)
\int^x dx_1 e^{-2q(x_1)\sqrt{\gamma}},
\label{2.8}   
\end{equation}
It follows from the canonical Poisson brackets of Liouville theory that 
\beq
 \Bigl \{ \vartheta(\sigma_1-i\tau),
\vartheta(\sigma_2-i \tau) \Bigr \}_{\hbox{PB}}=
2\pi   \delta'(\sigma_1-\sigma_2).
\label{a4.8} 
\eeq 
where $\vartheta=q'$. Doing the same construction with the functions $g$, one 
introduces another field $\bar \vartheta$ with opposite chirality. Altogether 
$\vartheta$ and $\bar \vartheta$ are the two chiral components of the two dimensional 
free field derived by the quantum B\"acklund transformation.  
Clearly the differential equation Eq.\ref{2.5} 
gives
$T=p^2+p'/\sqrt{\gamma}$, or equivalently
$T=(\vartheta')^2+\vartheta''/\sqrt{\gamma}$. The last expression coincides with
$U_1$ Sugawara stress-energy tensor with a linear  term. From
the
viewpoint of differential equations, these relations are simply
the well-known Riccati equations associated with the Schr\"odinger
equation Eq.\ref{2.5}. An easy computation shows that  
$T$ satisfies  the Poisson bracket Virasoro algebra with
$C_{\hbox{class}}=3/\gamma$. We shall consider the typical
situation of a
cylinder with $0\leq \sigma \leq \pi$, and $-\infty \leq \tau
\leq \infty$.  After appropriate coordinate
change, this may describe one handle of a surface with arbitrary genus. Then
$T$ is periodic in $\sigma$ with period $2\pi$, and the standard 
Virasoro generators are given by $L_n=\int _0^{2\pi} d\sigma e^{-in\sigma} T(\sigma-i\tau)$. 
Under the corresponding Poisson bracket action, it is easy to see that $f_1$, and $g_1$ 
are primary fields with weights $-1/2, 0$ and $0, -1/2$ respectively, so that 
$f_1^{-2}$ has weight $1,0$. Thus it is the classical version of the screening 
operator in accordance with Eq.\ref{2.5}. Classically, the Virasoro weights are simply additive, 
so that the weight of $\exp(-J\sqrt{\gamma}\Phi)=(\exp(-\Phi\sqrt{\gamma}/2))^{2J}$ is $-J,-J$. 
Therefore, we 
get $1,1$ for $J=-1$ in agreement with the fact that the potential term of the action 
Eq.\ref{c2.1} Is conformally invariant.
\subsection{Quantization (outline)} 
The parameter $\gamma$ plays the role of $\hbar$. In this operator method, one starts from a free 
quantum field  with chiral components  $\vartheta$ (and $\vartheta$) satisfying the quantum version 
of Eq.\ref{a4.8}, that is, one replaces $i$ times Poisson brackets by commutators. The above  
free-field expressions are extended to the quantum case by using  normal ordering and 
allowing for finite renormalizations of the classical constants. 
So we write $T\propto :\vartheta'^2:+Q\vartheta''$, which gives a central charge 
$C_L=1+3Q^2$. $C_L$ is the new free paremeter. The chiral vertex operatos are constructed from 
the quantum version of (powers of) Eq.\ref{2.8} with ``powers'' of $f_1$ defined by normal ordered 
exponentials of the type $:\exp(2J \alpha \vartheta):$, with $\alpha$ specified as follows. 
One imposes that the choice $J=-1$ again gives an operator of weight $1,0$, 
which is the screening operator. There are two choices, the 
two well known screening charges:  
\beq
\alpha_\pm=\sqrt{C_L-1\over 12}\pm \sqrt{C_L-25\over 12}. \quad  
\alpha_+\alpha_-=2. 
\label{screen}
\eeq     
With these choices, the fields with $J=1/2$ satisfy a quantum version of the 
Schr\"odinger equation \ref{2.5}, which is equivalent to the Virasoro null-field decoupling equation 
at level 2. Using its monodromy, and the associativity of the operator products ---or, more precisely 
the polynomial equations within the Moore Seiberg formalism\cite{MS}---, one 
deduces\cite{CGR,GS,GR1}  
 the OPA of the chiral vertex operators. A quantum group structure emerges of the type 
$U_q(sl(2))\odot U_{\qhat}(sl(2))$, with
\beq
q=e^{ih},\quad 
\qhat=e^{i\hhat}, \qquad 
h=\pi {\alpha_-^2\over 2},\quad 
\hhat=\pi {\alpha_+^2\over 2}. 
\label{qprms}
\eeq
The two quantum group parameters are related by 
\beq
h\hhat=\pi^2,\quad h+\hhat={C_L-13\over 6}.
\label{h-hhat}
\eeq
The first relation shows that they are, in a sense, dual pairs. The above symbol $\odot$ has
 a special   meaning which will be clarified as we go on. The Hilbert space of states is of course 
 a direct sum of Virasora Verma modules. They are characterized by the eigenvalue of the 
free field rescaled zero-mode momentum $\varpi$ defined by writing
\beq
\vartheta(x)=q_0+ip_0 x+i\sum_{n\not=0} e^{-inx} {p_n\over n},\quad
\varpi={2i\over \alpha_-}p_0. 
\label{p0def}
\eeq
A Verma module is charaterized by  the highest-weight eigenvalue 
\beq
\Delta(\varpi)={h \over 4\pi}(\varpi_0^2-\varpi^2), \quad 
\varpi_0=1+{\pi\over h}
\label{Delta}
\eeq
of $L_0$. For $\varpi=\pm \varpi_0$, $\Delta$ vanishes. This describes the two 
$Sl(2,C)$ invariant states. In general, the spectrum of $\varpi$ eigenvalues is of the form
\beq
\varpi_{J, \Jhat}=
\varpi_0+2\Je. 
\label{varpiJ}
\eeq
where $\Je$  which is called the effective spin specifies the  representations\footnote{
We refer to the original articles\cite{CGR,GS,GR1} 
 to explain  this point which is 
not central in these lectures. For recent developments in this direction 
see refs.\cite{CGS}.}
 of $U_q(sl(2))\odot U_{\qhat}(sl(2))$. If $\Je$ is generic, 
the corresponding Verma module is simply 
the bosonic Fock space of the free field $\vartheta$. If $\Je$ may be written as 
$\Je=J+\pish \Jhat$, with $2J$ $2\Jhat$ non negative integers,  are 
 this Fock space involves null states and the irreducible 
Verma modules must be projected out. We shall assume that this is done if needed. 
In this latter case $J$ and $\Jhat$ are ordinary quantum group spins.  
\subsection{Chiral operator algebra}
 For continuous spins, $J$ and $\Jhat$ loose 
meaning in general, and only $\Je$ makes sense. However, 
in our strong coupling discussion---where $\hspi$ is not rational---we 
shall only need to deal with the case where $\Je$ takes the form $J+\Jhat\pish$, with
   $J$ $\Jhat$ 
rational numbers (not necessarily positive integers). Then $J$ and  $\Jhat$ may be 
defined uniquely from $\Je$, so that we use them in our discussion. 
We shall try to display the structure of the result as imply as possible by drawing pictures. 
There will be   a double line for each  Verma module, with $J$ and $\Jhat$   
    represented by  a solid and  a dashed line, respectively, 
  Concerning  the three-leg conformal blocks, the matrix elements of the  
chiral vertex operators (which are the chiral components of the 
Liouville exponentials)  denoted\footnote{
The tilde is to distinguish from amother normalisation 
(see ref.\cite{CGR}).}
 by $\Vt$  are represented by  three leg vertices.  
The operator product algebra (OPA) is thus pictured by diagrams with 
basic elements
\beq
\fig{10.5cm}{2.5cm} {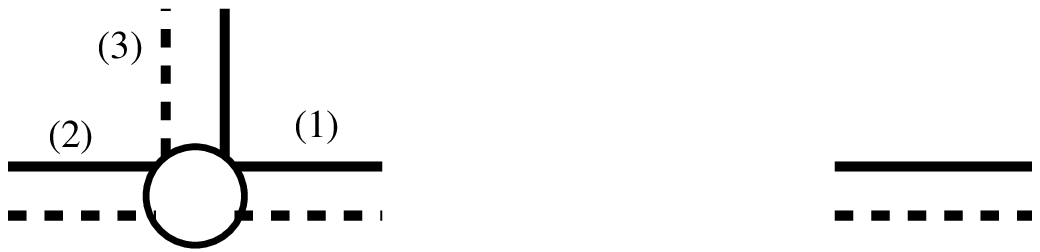} 
\label{cb|elemV}
\eeq
which correspond to
\beq
<\J{2}\{\beta_2\}|\Vt^{\J{1},\{\beta_1\}}(z)\J{3},\{\beta_3\}>, \quad 
\sum_{\{\beta\}}{|\Je, \{\beta\}><\Je, \{\beta'\}|\over 
<\Je, \{\beta'\}|\Je, \{\beta\}>}.
\label{elements}
\eeq
From now on we use the world sheet variable $z=\exp(ix_+)$. 
The symbol  $\{\beta\}$ stands for  a multi-index that characterizes the 
descendent of the Verma module---or the primary field. 
In order to avoid clumsy drawings, we omit it from the diagrams. It is understood that it is 
summed over for each intermediate (double) line, as shown on the 
last  equation above. 
 On the 
contrary, {\bf for the effective spins, the summation over intermediate 
line should be performed only when 
indicated}. Operatorially, this reflects the fact that the $V^{(\Je)}$ operators 
possess an additional index which specifies the shift of $\Je$ between bra and ket. 
More precisely, for continuous $\Je$, one defines operators $\Vt^{(\Je),\{\beta\}}_{\me}$ 
with the condition that (${\cal Z}$ represents the set of non negative integers)  
\beq
\Je+\me =\nu +\pish \nuhat,\quad \nu,\> \nuhat \in {\cal Z}, 	
\label{screen1}
\eeq
such that
\beq
<\J{2} \{\beta_2\}|\Vt^{\J{1},\{\beta_1\}}_{\m{1}}(z)|\J{3},\{\beta_3\}>=0, 
\> \hbox{unless}\> \m{1}=\J{3}-\J{2}
\label{element2}
\eeq 
 
 For instance, a four leg conformal block 
is drawn as 
\beq
\fig{4.8cm}{2cm}{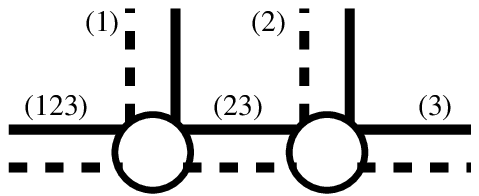} \up{\Leftrightarrow 
<\J{123}, \{\beta_{123}\}|\Vt_{\J{23}-\J{123}}^{\J{1},\{\beta_1\}}(z_1) 
\Vt_{\J{3}-\J{23}}^{\J{2},\{\beta_2\}}(z_2)|\J{3},\{\beta_3\}>} 
\label{exemple}
\eeq
These chiral vertex operator are closed by fusion 
and braiding. Let us illustrate this in the 
particular case where all the $\Jhat$'s are zero. Using therefore simple lines one 
has the following representation. For fusion, 
\beq
\fig{5.5cm}{2.5cm}{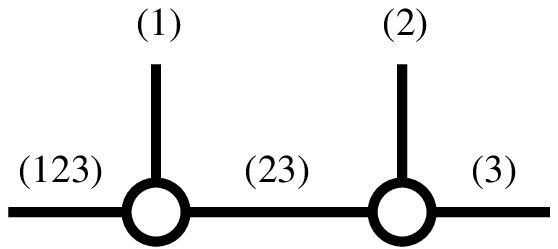} {\displaystyle =\sum_{J_{12}} \>F_{{J_{23}},{J_{12}}}\!\!\left[^{J_1}_{J_{123}}
\,^{J_2}_{J_3}\right]\> \times 
\atop\phantom{\hbox {FUSION MATRIX}} } \fig{3cm}{3.5cm}{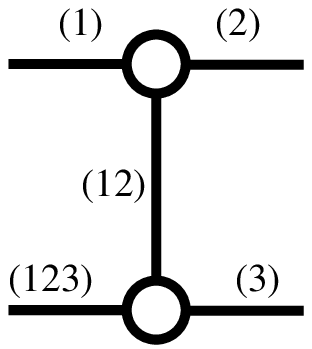}
\label{cb|sV,uV}
\eeq
where the fusion matrix is given by  
\beq
F_{{J_{23}},{J_{12}}}\!\!\left[^{J_1}_{J_{123}}
\,^{J_2}_{J_3}\right]
=
\left\{ ^{J_1}_{J_3}\,^{J_2}_{J_{123}}
\right. \left |^{J_{12}}_{J_{23}}\right\}_q.
\label{fus}
\eeq
On the right there appears    the q 6j symbol.   
 Similarly, the braiding is represented by 
\beq
\fig{5.5cm}{2.5cm}{sV.eps} {\displaystyle =\sum_{J_{13}} \>
B^\pm_{{J_{23}},{J_{13}}}\!\!\left[^{J_1}_{J_{123}}
\,^{J_2}_{J_3}\right]\>  
\atop\phantom{\hbox {FUSION MATRIX}} } \fig{5.5cm}{2.5cm}{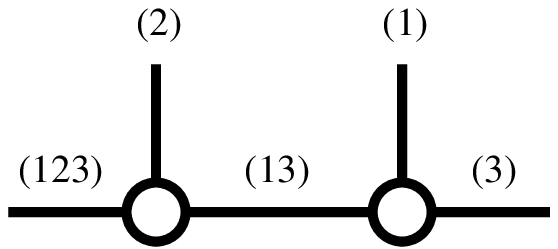}
\label{cb|sV,tV}
\eeq
where the braiding matrix is related to the above by 
\beq
B^{\pm}_{{J_{23}},{J_{12}}}\!\!\left[^{J_1}_{J_{123}} 
\,^{J_3}_{J_2}\right] 
=e^{\pm i \pi \left( \Delta_{J_{123}}+\Delta_{J_2}-\Delta_{J_{23}}
-\Delta_{J_{12}}\right )}
F_{{J_{23}},{J_{12}}}\!\!\left[^{J_1}_{J_{123}} 
\,^{J_2}_{J_3}\right].  
\label{braid}
\eeq 
In the general case the structure is similar, with the corresponding extended 6j symbols 
associated with our $U_q(sl(2))\odot U_{\qhat}(sl(2))$, which are given by 
\beq
F_{{\J{23}},{\J{12}}}\!\!\left[^{\J{1}}_{\J{123}}
\,^{\J{2}}_{\J{3}}\right]
=
\left\{\left\{ ^{\J{1}}_{\J{3}}\,^{\J{2}}_{\J{123}}
\left. \right |^{\J{12}}_{\J{23}}\right\}\right\}
\left\{\left\{ ^{\Jehat_1}_{\Jehat_3}\,^{\Jehat_2}_{\Jehat_{123}}
\bverthat\bigr. \, ^{\Jehat_{12}}_{\Jehat_{23}}\right\}\right\}_{\qhat}.
\label{Fjjhat}
\eeq   
The double brace means 
6j symbols with shifted entries (see ref.\cite{GR1} for details). The symbol  $\Jehat$ 
is defined  by $\Jehat=\Jhat +\hspi J\equiv \hspi \Je$.  
The normalization of
the chiral vertex operators $\Vt$ is fixed by the condition that
their fusion and braiding  matrices be exactly equal to q 6j symbols---not
simply proportional as they would be with other choices. Then, their
 highest weight matrix element is given by
\beq
<\J{2}| \Vt^{(\J{1})}(z) |\J{3}>=
g_{\J{1}, \J{3}}^{\J{2}}
\label{VtVEV}
\eeq
where the $g$'s, which are 
 called coupling constant,  are not simply determined by the
quantum group symmetry---they are not trigonometrical functions---but
from the monodromy properties of the quantum version of Eq.\ref{2.5}, the
null vector decoupling equation.
\subsection{The weak coupling regime} 
\subsubsection{The Liouville exponentials}
We have to put the two chiralities together. 
The $x_-$ components---which we did not 
consider so far---are of course similar to the above. 
Their  quantum numbers will be  distinguished by 
a bar, so we have $\varpib$, $\Jb$, $\Jbhat$ and so on. Graphically, the Verma modules 
are represented by gray lines in contrast with the black ones of the $x_+$ components.   
The Liouville exponentials may be determined as follows: one writes 
  general chiral decompositions 
and imposes that they be local, with $\sigma$ and $\tau$ being space and time coordinates 
respectively. Remarkably, this determines them completely, and the result may be 
consistently restricted to Verma modules 
such that 
\beq
J=\Jb,\quad \Jhat =\Jbhat
\label{windconf}
\eeq 
To represent this restriction, we introduce springs that link our lines as follows 
\beq
\centre{\fig{4cm}{3.2cm}{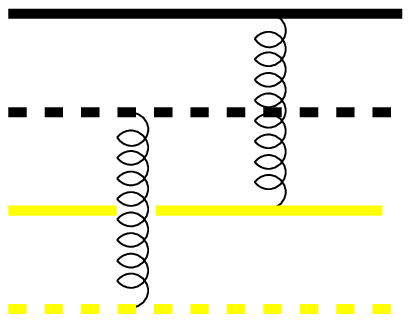}{}} \begin{array}{c} 
J\nnn\phantom{J}\nnn\Jhat\nnn\phantom{J}
\nnn\Jb=J\nnn\phantom{J}\nnn\Jbhat=\Jhat
\nnn
\end{array}
\label{wcond}
\eeq
Then the Liouville three point function  is  represented by 
\beq
\centre{\fig{4cm}{2.8cm}{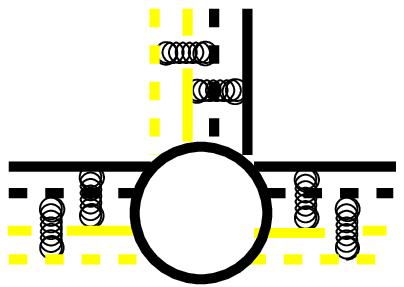}}; \>\hbox{such that}\>  
\centre{\fig{4cm}{2.8cm}{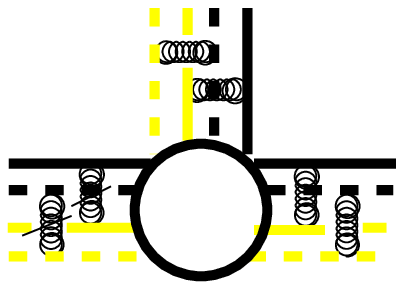}{}} =0
\label{Liouvexp}
\eeq
The right-hand side drawing means that the Liouville exponentials 
applied to a state satisfying condition Eq.\ref{windconf} 
only give states satisfying the same 
condition. Operatorially, the Liouville exponentials are given  by 
\beq
e^{\textstyle -\Je\alpha_-\Phi(z, \zb )}=
\sum_{\me} \Vt^{(\Je)}_{\me}(z) \Vbt^{(\Jeb)}_{\meb}(\zb). 
\label{Liouvop}
\eeq
This chiral decomposition only involves $\Vt$ and $\Vbt$ operators with the 
same $\Je$ and $\me$, in accordance with the diagrams just written 
(recall Eq.\ref{element2}). It is the quantum version of the 
classical expression 
Eq.\ref{2.4} and of its higher spin generalizations.    
The  properties of the quantum Liouville exponentials  under fusion and braiding are deduced from 
the chiral ones summarized above, and are consequences 
 of the orthogonality of 
the q 6j symbols. For the standard ones it reads 
\beq
\sum_{J_{23}}
\left\{ ^{J_1 }_{J_3 }
\> ^{J_2 } _{ J_{123}}
\right.
\left |^{ J_{12}}
_{J_{23}}\right\}
\left\{ ^{J_1 }_{J_3 }
\> ^{J_2 } _{ J_{123}}
\right.
\left |^{ K_{12}}
_{J_{23}}\right\} =\delta_{J_{12}-K_{12}}. 
\label{orth}
\eeq
As a result, the braiding matrix of the Liouville exponentials are 
equal to one, so  
that they are mutally local, i.e.
$$
\up{\sum_{J_{23}}}
\> \fig{6.4cm}{2.7cm}{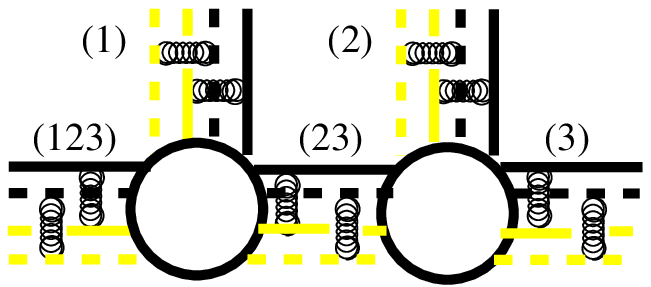} 
$$
\beq
\up{=\sum_{J_{13}}} 
\> \fig{6.4cm}{2.7cm}{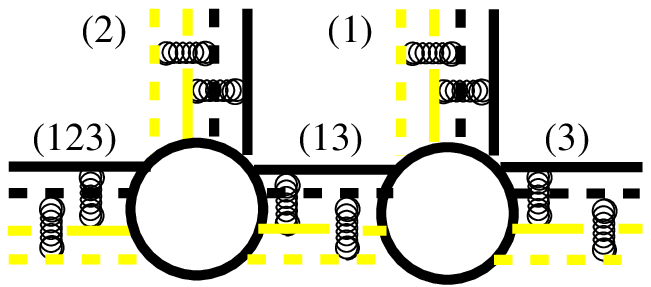} .
\label{Liouvt}
\eeq
They are moreover  closed by fusion, following the diagram 
$$  
\up{\sum_{J_{23}}}
\>\fig{6.4cm}{2.7cm}{Liouvs.eps} 
\up{=\sum_{J_{12}}} 
\>\fig{4cm}{4.3cm}{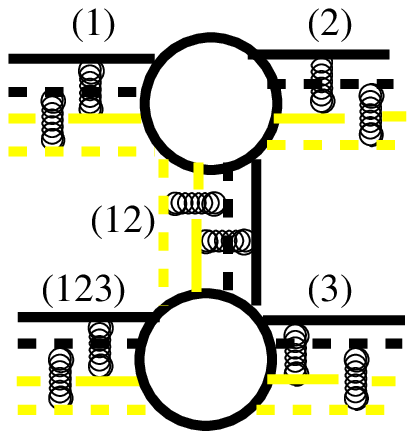} 
$$    
As usual, the upper vertex  on the right hand side is a book keeping device
for descendents. To leading order, at short distance, the coefficient 
is determined by its expectation value between highest weight states.  
According to Eqs.\ref{VtVEV}\ref{Liouvop} it 
follows that
\beq
<\J{2}| e^{\textstyle -\J{1}\alpha_-\Phi(z, \zb )} |\J{3}>=
|g_{\J{1}, \J{3}}^{\J{2}}|^2. 
\label{LiouvVEV}
\eeq
and   the contribution of $\exp(-J_{12}\alpha_-\Phi)$---and of its 
descendent---in the operator 
product of $\exp(-\J{1}\alpha_-\Phi)$ with  $\exp(-\J{2}\alpha_-\Phi)$,  is 
proportional to $|g_{\J{1} \J{2}}^{\J{12}}|^2$. Thus the coupling constants 
determine the three point functions of the OPA, as expected.
Concerning the braiding diagram, it explicitly means that
\beq
e^{\textstyle -\J{1}\alpha_-\Phi(z_1, \zb_1 )}
e^{\textstyle -\J{2}\alpha_-\Phi(z_2,\zb_2)}=
e^{\textstyle -\J{2}\alpha_-\Phi(z_2,\zb_2)}
e^{\textstyle -\J{1}\alpha_-\Phi(z_1, \zb_1 )},
\label{loc}
\eeq
where we are are equal $\tau$, so that $z_1=\exp{(\tau+i\sigma_1)}$,
$z_2=\exp{(\tau+i\sigma_2)}$.

In general, applying a similar discussions to the higher point functions, one generates a 
set of diagrams where condition Eq.\ref{windconf} is obeyed on every line. Since this has 
an obvious analogy with quark diagrams, we may say that chirality is confined.   
\subsubsection{The coupling to matter} 
One  represents\cite{G5}  matter by another copy of the
theory summarized above, now with central charge $c=26-C_L$. 
 One constructs local
fields in analogy with Eq.\ref{Liouvop}:
\beq
e^{\textstyle -(J\alpha'_-+\Jhat\alpha'_+)\Phi'(z, \zb )}=
\sum _{m,\, \mhat}
\Vt'\,^{(J,\,\Jhat)}_{m\,\mhat}(z)\,
\Vbt'\,^{(J,\,\Jhat)}_{m\,\mhat}(\zb).  \>
\label{L'expdef}
\eeq
Symbols pertaining to matter are distinguished by a prime. 
In particular $\Phi'(z, \zb )$ is the matter field  (it
commutes with $\Phi(z, \zb )$), and $\alpha'_\pm$ are the matter
screening charges
\beq
\alpha'_\pm=\mp i\alpha_\mp.
\label{alpha'def}
\eeq
The correct dressing of these operators by gravity is
achieved by considering the vertex
operators
\beq
{\cal W}^{J,\Jhat}\equiv
e^{\textstyle -((-\Jhat-1)\alpha_-+J\alpha_+)\Phi
 -(J\alpha'_-+\Jhat\alpha'_+)\Phi'}
\label{vertexdef}
\eeq
which is an operator of weights $\Delta=\bar \Delta=1$.
In particular for $J=\Jhat=0$, we get the cosmological term
$\exp (\alpha_-\Phi)$. The three-point function was computed
in refs\cite{G5,GR1}. The corresponding product of coupling
constants gives the correct leg factors after drastic
simplifications.
Of course, the dressing is such that  the Verma modules  involved 
satisfy the appropriate BRST condition. Graphically, 
they may be represented as  
\beq
\fig{4.6cm}{4.4cm}{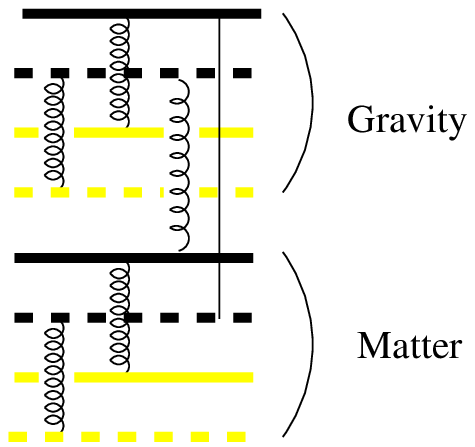} 
\label{cb|wcond2}
\eeq
In this drawing the  straight line between the first and sixth  lines  
indicates that the $J$'s satisfy the condition $J+\Jhat'+1=0$, as required by 
the dressing condition.  

\subsection{The strong coupling regime} 
\subsubsection{The barrier}
Eqs.\ref{screen} for the screening charges are real for 
$C_L>25$, pure imaginary for $C_L<1$ and complex otherwise. The weights 
of the primary fields with spins $J$ $\Jhat$ are given by Kac's formula
\beq
\Delta(J,\, \Jhat) =
{C_L-1\over 24} -{1\over 24}
\left [(J+\Jhat+1)\sqrt{C_L-1} -(J-\Jhat)\sqrt{C_L-25}\right ]^2, 
\label{kac}
\eeq
which, for real $J$ $\Jhat$, is real for $C_L>25$, and $C_L<1$, and complex 
in general otherwise.  In the zero coupling limit  $\gamma\to 0$, 
$C_L\sim 3/\gamma$ blows up. Thus the region $C_L>25$ is called the 
weak coupling regime of gravity, which is connected with the classical limit 
$\gamma\to 0$. There the above formula is real and the construction 
of the Liouville exponentials just summarized applies. 

Can we go beyond $C_L=25$?. At this moment it seems that one cannot make sense
of operators with complex $\Delta$. For $1<C_L<25$, Eq.\ref{kac} still gives 
real weights in two particular cases. First $\Delta$ is real and negative 
if $J=\Jhat$.  In the type of picture we are using (see \ref{wcond}), 
this means that  a spring is introduced between the  two lines, but here 
they have   the same 
chirality. Second, $\Delta$ is real and positive if $J+\Jhat+1=0$. This 
kind  of condition (a sort of repulsive ``interaction'')  already appeared 
on the diagram \ref{cb|wcond2}. It is  
represented by a solid line. Thus the subspaces with real $\Delta$'s  
are represented graphically as follows. 
\beqa
\centre{\fig{4.5cm}{1.2cm}{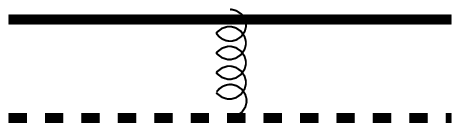}{}}\> 
&\leftrightarrow& \> \Jhat=J,\quad \Delta(J,J)\>  \hbox{real}\>  
<0,  \nnn
&\phantom{\leftrightarrow}&\nnn
\centre{\fig{4.5cm}{1.2cm}{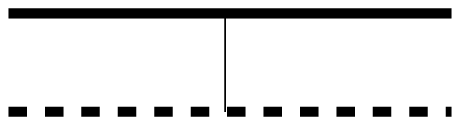}{}}\> 
&\leftrightarrow& \> \Jhat=-J-1,\quad \Delta(J,-J-1)\>  \hbox{real}\> 
>0. 
\label{real}
\eeqa
Now we return to the Liouville exponential. Does it respect this 
reality conditions? The answer is no, since, in Eq.\ref{Liouvop}, one sums over 
$\me$ and $\meb$ independently, so that the shift of $\Je$ and $\Jeb$ are not 
corrolated.  Graphically, this is illustrated as 
follows 
\beq
\centre{\fig{4.3cm}{2.9cm}{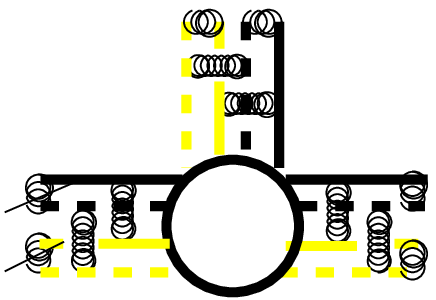}{}} \not =0,  
\label{clash}
\eeq
and by a similar relation with the other condition.
Operiatorially, this means that we may try to use Liouville exponentials 
with real weights (depicted by the additional springs on the vertical 
lines) but applied to states with real weights they give states with 
complex weights so that they loose meaning in the strong coupling regime. 
\subsubsection{The new set of chiral fields} 
The solution to this problem is as follows. The chiral operator algebra 
of the $V$ fields has a natural extension to  $C_L<25$, albeit with 
complex weights. One constructs  different types of local 
operators, where the conditions $\Jb=J$ and $\Jbhat=\Jhat$ 
are not imposed. They are built from the same chiral vertex operators 
as the Liouville exponentials, but one takes  different  combinations 
so that they respect one of the two reality conditions Eq.\ref{real}. 
Following our  graphical rules, their three point function are of the type 
$$
\centre{\fig{4cm}{2.7cm}{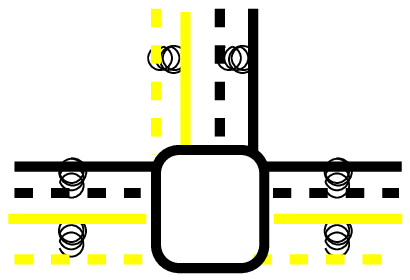}{}} \>\hbox{or}\>
\centre{\fig{4cm}{2.7cm}{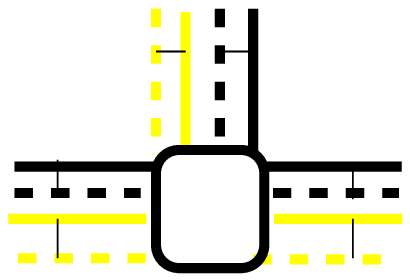}{}} 
$$
The reality 
conditions  on the Verma modules  do  not link the two chiralities any 
longer, so we may say that chirality is deconfined.  The possibility of 
 constructing these  operators consistently  is based on the truncation theorems that hold 
for special values of $C_L$, i.e. $C_L=7$, $13$, $19$. One first constructs 
chiral operators which are closed by fusion and braiding. 
They are particular 
linear combinations of the chiral vertex operators  drawn on 
figure \ref{cb|elemV} These so called called  physical operators $\chi_\pm^{(J)}$   
 have three point functions of the type (for the $x_+$ components)
$$
\centre{\fig{3.3cm}{2.7cm}{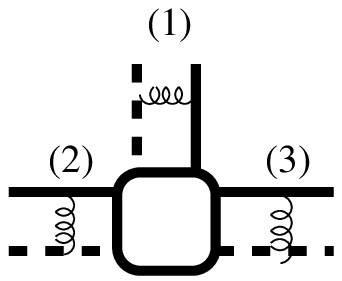}{}} \>\hbox{and}\>
\centre{\fig{3.3cm}{2.7cm}{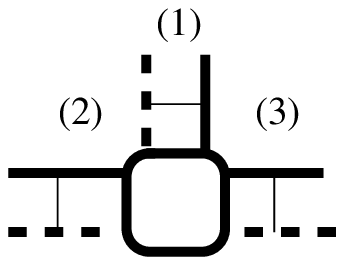}{}} 
$$
which represent 
\beq
<\J{2}^\pm\{\beta_2\}|\chi_\pm^{\J{1}^\pm,\{\beta_1\}}(z)
|\J{3}^\pm,\{\beta_3\}>, 
\label{chiop}
\eeq 
where 
\beq
\Je^\pm=J(1\mp \pish)+{\pi \over 2h}(1\pm 1)
\label{Jpm}
\eeq 
A similar representation with gray lines holds of course for the 
$x_-$ components. We do not discuss it explicitly. 

Now is a good time to recall
 the truncation theorems
which hold for
\beq
C=1+6(s+2),\quad s=0,\pm 1.
\label{Cspecdef}
\eeq
 First  define the physical
Hilbert space
\beq
{\cal H}_{ \hbox{\footnotesize phys}}^{\pm}\equiv
   \bigoplus_{r=0}^{1\mp s}   \bigoplus_{n=-\infty}^\infty
{\cal H}^\pm_{r/2(2\mp s)+n/2}
\label{Hphysdef}
\eeq
where  ${\cal H}^\pm_{\Je}$  denotes the Verma modules with highest
weights $|\Je^\pm>$.
 The tree leg conformal bloch of the  $\chi$  operators 
take the  
form
$$  
\fig{3.4cm}{2.5cm}{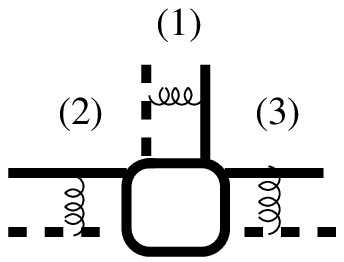}
\up{=(-1)^{(2-s)(2J_3+\nu(\nu+1)/2)}}
\fig{3.4cm}{2.6cm}{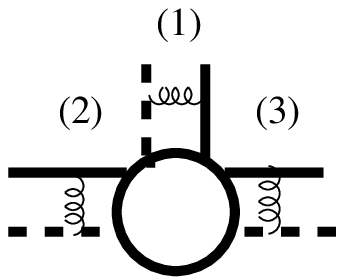}
$$
$$  
\fig{3.4cm}{2.5cm}{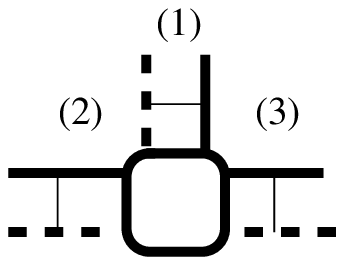}
\up{=(-1)^{(2+s)(2J_3+\nu(\nu+1)/2)}}
\fig{3.4cm}{2.6cm}{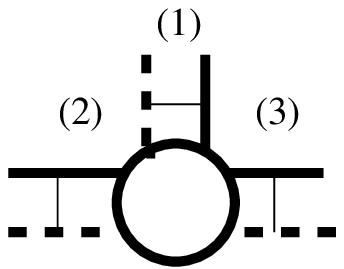}
$$
if $\nu=J_1+J_3-J_2\in {\cal Z}_+$, and $2J_i \in {\cal Z}/(2\mp s)$; 
and they are taken to vanish otherwise. 
By construction the physical operators are restricted to 
${\cal H}^\pm_{ \hbox{\footnotesize phys}}$. 
Denote their ensemble by ${\cal A}^\pm_{\hbox{\footnotesize
phys}}$. The basic
properties  of the special values Eq.\ref{Cspecdef} is the

\noindent TRUNCATION THEOREM:

 For $C_L=1+6(s+2)$, $ s=0$, $\pm 1$,
the above set ${\cal A}^+_{\hbox{\footnotesize
phys}}$ (resp. ${\cal A}^-_{\hbox{\footnotesize
phys}}$) is closed by braiding and fusion. 

Let us display, for instance the braiding properties. One has 
$$
\up{\sum_{J_{23}}}  \fig{4.3cm}{1.6cm}{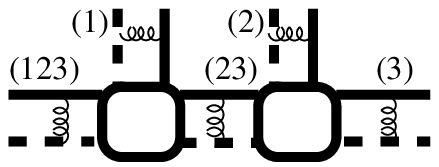}\up{\> =
\sum_{J_{13}}  (-1)^{\pm 2i\pi (2-s)J_1J_2} \>} \fig{4.3cm}{1.6cm}{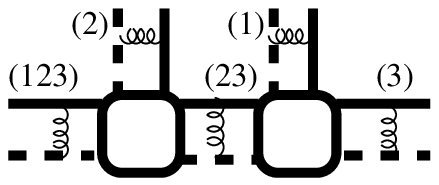}
$$
$$
\up{\sum_{J_{23}}}  \fig{4.3cm}{1.6cm}{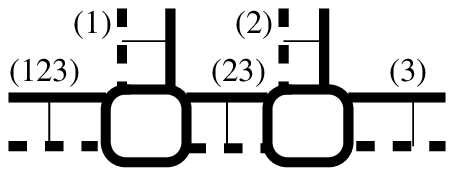}\up{\> =
\sum_{J_{13}}  (-1)^{\pm 2i\pi (2+s)J_1J_2} \>} \fig{4.3cm}{1.6cm}{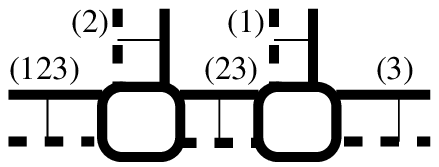}
$$
Note an important difference with the braiding properties of 
the previous chiral components themselves (figure \ref{cb|sV,tV}). 
On the left hand side of the last two drawings one has to  sum over the 
intermediate $J_{23}$, contrary to \ref{cb|sV,tV}. Thus the $\chi$ operators, 
contrary to the $\Vt$ do not have a quantum number specifying the 
shift of $\Je$. As a matter of fact, the operatorial relation between 
$\chi$ and $\Vt$ operators consistent with the three and four leg diagrams just 
displayed is as follows.     
\beq
\chi^{(J_1)}_\pm\>
{\cal P}_{{\cal H^\pm}_{J_{3}}}
=\sum_{\nu \equiv J_1+m\in \cal Z_+}
(-1)^{(2\mp s)(2J_3+\nu(\nu+1)/2)} 
\Vt_{\me^\pm}^{(\J{1}^\pm)} \>{\cal P}_{{\cal H}^\pm_{J_3}},
\label{chidef}
\eeq
where ${\cal P}_{{\cal H}^\pm_J}$ is the projector
on ${\cal H}^\pm_{J}$, and $\me^\pm=m(1\pm \pish)$. Note that, 
on the contrary, 
the operators $V^{(\Je^\pm)}_\me$ themselves   cannot be consistently restricted 
to either ${\cal H}^+_{\hbox{\footnotesize 
phys}}$ or ${\cal H}^+_{\hbox{\footnotesize 
phys}}$. 
Operatorially, the last two figures correspond to 
\beq
\chi^{(J_1)}_\pm \chi^{(J_2)}_\pm =
e^{2i\pi \epsilon (2\mp s)J_1J_2 }
\chi^{(J_2)}_\pm \chi^{(J_1)}_\pm,
\label{chibraid}
\eeq
where $\epsilon=\pm 1$ is fixed by  the ordering of the operator
on the left-hand side in the usual way.    
Next we construct local fields out of the chi  fields.  The braiding 
of the chi fields is a simple phase. From the spectrum of
the $J$'s, it follows that this  phase factor is of the form
$\exp(i\pi N/2(2\mp s))$, where $N\in \cal Z$. Thus, we have
parafermions.  As shown in
ref.\cite{GR1},
 simple products  of the form
$\chi^{(J)}_\pm
\chib^{(\Jb)}_\pm$, with $J-\Jb \in \cal Z$  are local.
 In such a product, the summations over
$m$, and $\mb$ are independent, while the summations over
$m$, $\mhat$, and $\mb$, $\mhatb$ are correlated. Now  we have a
complete reversal of the
weak coupling situation summarized by the drawing \ref{Liouvexp} :the new
fields preserve the reality condition, but {\bf do not preserve
the equality between $J$ and $\Jb$ quantum numbers}.  Thus, as
 stressed in ref.\cite{GR1}, in the strong coupling
regime  we observe a sort of deconfinement of chirality. 

\section{The Liouville string}
One may consider two different problems. First, one may build a
full-fledged string theory, by coupling, for instance,
the above  with
$26-C_L$ free  
fields $\vec X$. A  typical string vertex is
of the form  
$\exp(i\vec k.\vec X) \chi^{(J)}_+
\chib^{(\Jb)}_+$,
where $\vec k$, $J$, and $\Jb$ are related so that this is
a $1,1$ operator. Here obviously, the restriction to real weight
is instrumental. Moreover, since one  wants the representation of
Virasoro algebra to be unitary, one  only uses the chi+ fields.
This line was already persued with noticable success in
refs.\cite{BG}. However, the N-point functions seem to be beyond
reach at present. Second a simpler problem seems to be tractable,
namely, we may proceed as in the construction of topological
models just recalled. We consider another copy of the present
strongly coupled theory, with central charge $c=26-C_L$. Since
this gives $c=1+6(-s+2)$, we are also at the special values, and
the truncation theorems applies to matter as well. This ``string
theory'' has no transverse degree of  freedom, and is thus
topological.
The complete dressed vertex operator is now
\beq
\vertex J, \Jb,
=
\chi_+^{(J)} \chib_+^{(\Jb)}\>
\chi'\, ^{(J)}_-  \chib'\, ^{(\Jb)}_-
\label{VertexSdef}
\eeq
As in the weak coupling formula, operators relative to
matter are distinguished by a prime.
The definition of the $\chib$ is similar to the above, with an
important difference. Clearly, the definition of
$\chi_+$ is not symmetric between $\alpha_+$, and
$\alpha_-$. The truncation theorems also holds if we
interchange the two screening charges. We re-establish some
symmetry between them by taking the other possible definition for
$\chib_+$.
Our results will then be  invariant by complex conjugation
provided we exchange $J$'s and $\Jb$'s. Thus left and right movers
are interchanged, which seems  to be a sensible requirement.
The spectrum of Verma modules involved may be depicted graphically 
as follows:
\beq
\fig{4.6cm}{4.4cm}{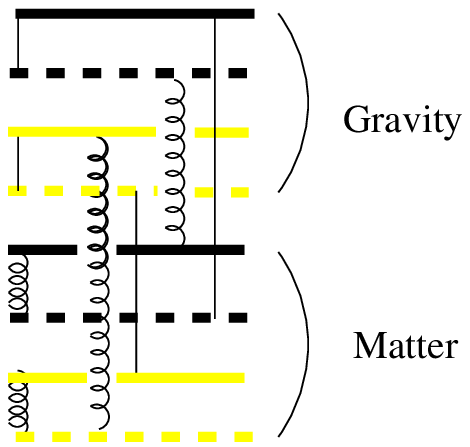}
\label{scond}
\eeq 
It is suggestive to compare with the corresponding drawing 
\ref{cb|wcond2}of the 
weak coupling regime. The vertical links between the lines are of the 
same nature, but they are distributed differently. A basic difference 
here is that  the present drawing is not  connected  since there are no 
links between the two chiral components. Thus,  chirality is deconfined. 
For $J=\Jb=0$, we get the new cosmological term
\beq
\vertex 0, 0, =\chi_+^{(0)}(z) \chib_+^{(0)}(\zb).
\label{cosmodef}
\eeq
Thus the area element of the strong coupling regime is
$\chi_+^{(0)}(z) \chib_+^{(0)}(\zb) dz d\zb$. It is factorized
into
a simple product of a single $z$ component by a $\zb$ component.
{}From this expression one may compute the string susceptibility
using the operator version of the DDK argument developed in
ref.\cite{G5} for the weak coupling regime. We refer to 
refs.\cite{GR2,GR3} for 
details. One finds 
\beq
\gamma_{\hbox {\scriptsize   str}}=(2-s)/ 2.
\label{1.17}
\eeq
The result is real for $c>1$ ($C_L<25$), contrary to the
continuation of the
weak-coupling equation
$\gamma_{\hbox {\scriptsize   str}}=2-Q/\alpha_-$.
Explicitly one has
\beq
\left \{
\begin{array}{rccc}
s& c & C_L & \gamma_{\hbox {\scriptsize   str}} \nnn
1& 7  & 19 & 1/2 \nnn
0& 13 & 13 & 1 \nnn
-1 & 19 & 7 & 3/2 \nnn
\end{array} \right.,\qquad
\left \{
\begin{array}{rccc}
s& c & C_L & \gamma_{\hbox {\scriptsize   str}} \nnn
2& 1  & 25 & 0 \nnn
-2& 25  & 1 & 2
\end{array} \right.
\label{1.19}
\eeq
The last two are the extreme points of the strong coupling
regime. The values at $c=1$, and $c=25$
 agree with the weak-coupling formula. The result is always
positive,
contrary to the weak-coupling regime. At $c=7$, we find the value
$\gamma_{\hbox {\scriptsize   str}}=1/2$ of branched polymers. 

Finally, we have computed  the N-point functions  with one
incoming and N-1 outgoing legs, defined as follows.
First
in general\cite{CGR} the two-point function of two $\Vt$
fields with spins $J_1,\, \Jb_1$, and $J_2,\, \Jb_2$
 vanishes unless $J_1+J_2+1=0$, and $\Jb+\Jb_2+1=0$. Thus
conjugation involves the transformation $J\to -J-1$.
Taking account of the exchange between   $J$ and $\Jhat$,
due to complex conjugation,
yields the following vertex operator for
the conjugate representation:
\beq
\vertexc J, \Jb,
=
\chi_+^{(J)}  \> \chib_+^{(\Jb)} \>
\chi'\,^{(-J-1)}_- \>\chib'\,^{(-\Jb-1)}_-.
\eeq
Thus we have computed the matrix elements
$\left <
\vertexc J_1, \Jb_1,
\vertex J_2, \Jb_2 ,
...
\vertex J_N, \Jb_N,
\right >$. The method is similar to the one developed in ref.\cite{dFK},
with a reshuffling of quantum numbers. In the weak coupling regime,
left and right quantum numbers are kept equal, while the ones  associated
with different  screening charge are chosen independently. In the strong
coupling regime, the situation is reversed: the reality condition
ties the quantum numbers which  differ by the   screening charge, but
the  quantum numbers with different chiralities  become independent.
See refs.\cite{GR2,GR3} for details.

\section{Concluding remarks}
We should probably stress that   no total  conformal spin has been introduced.
Indeed, although, the conformal spin is non zero for the gravity, and matter components
of our vertex operators separately,
 the  total weights 
for the  left and right components are kept equal to one.

One may wonder why the present approach succeeds to break through the $c=1$
barrier, in sharp contrast with the other ones. This may be traced to
the fact that
we first deal with the chiral components of gravity and matter separately before
reconstructing the vertex operators. This is more painful than the matrix model
approach which directly constructs the expectation values of the dressed matter
operators. However, in this way we have a handle over the way the gravity quantum
numbers are coupled, and so we may build up vertex operators which change the
gravity chirality. This seems to be the key to the $c=1$ problem, since this
quantum number plays the role of order parameter.

Let us turn to a final  remark. The redefinition of the
cosmological term led us to modify the KPZ
formula\cite{KPZ}. On the other hand, in standard studies
of the matrix models  or  KP flows, one first derives
$\gamma_{\hbox {\scriptsize   str}}$ and deduces the
value of the central charge
by  assuming that the KPZ formula holds. In this
way of thinking, one would start from our formula
Eq.\ref{1.17} and apply KPZ, which would lead to
a different value of the central charge, say $d$. It is
easy to see that for $c=1+6(-s+2)$ one gets
$d=1-6(2-s)^2/2s$. This is the value of
a $2, s$ minimal model! What happens is that
in terms of $d$, we have $\gamma_{\hbox {\scriptsize
str}}=(d-1+\sqrt{(d-1)(d-25)})/12$, in contrast with the
KPZ formula $(d-1-\sqrt{(d-1)(d-25)})/12$. Thus the strongly coupled
topological theories  may be given by another branch of $d<1$
theories.

\end{document}